\documentclass[12pt]{article}
\usepackage{amsmath, amssymb}
\usepackage{graphicx,psfrag, subcaption}
\usepackage{enumerate}
\usepackage{natbib}
\usepackage{bbm}
\usepackage{url} 

\newcommand{\blind}{0}

\newcommand{\Lnet}{\mathfrak{L}}
\newcommand*{\smallG}{\scalebox{0.54}{$\mathcal{G}$}}
\newcommand{\mrf}{\stackrel{\smallG}{\sim}}
\newcommand{\bnr}{\stackrel{\smallG}{\longrightarrow}}

\DeclareMathOperator{\nach}{ne}

\DeclareMathOperator{\pa}{pa}

\DeclareMathOperator{\child}{ch}
\DeclareMathOperator{\family}{fam}

\DeclareMathOperator{\cop}{co-pa}
\DeclareMathOperator{\dg}{deg_{\mathcal{G}}}

\begin{document}

\def\spacingset#1{\renewcommand{\baselinestretch}%
{#1}\small\normalsize} \spacingset{1}


\if0\blind
{
  \title{\bf  Point patterns occurring on complex structures in space and space-time: An alternative network approach}
  \author{Matthias Eckardt\thanks{Financial support from the Spanish Ministry of Economy and Competitiveness via grant MTM 2013-43917-P is gratefully acknowledged.}\hspace{.2cm}\\
    Department of Computer Science, Humboldt Universit\"{a}t zu Berlin, Berlin, Germany\\
    and \\
   Jorge Mateu\\
    Department of Mathematics, University Jaume I, Castell\'{o}n, Spain}
  \maketitle
} \fi

\if1\blind
{
  \bigskip
  \bigskip
  \bigskip
  \begin{center}
    {\LARGE\bf Title}
\end{center}
  \medskip
} \fi

\bigskip
\begin{abstract}
This paper presents an alternative approach of analyzing possibly multitype point patterns in space and space-time that occur on network structures, and introduces several different graph-related intensity measures. The proposed formalism allows to control for processes on undirected, directional as well as partially directed network structures and is not restricted to linearity or circularity.   
\end{abstract}

\noindent%
\vfill

\newpage
\spacingset{1.45} 

\section{Introduction}\label{sec:int}
The last decade witnessed an extraordinary increase in interest in  the analysis of network related data within numerous disciplines. This pervasive interest is on the one hand caused by a strongly expanded availability of network data. Examples of such network data are social network, traffic networks or genetic networks. On the other hand, underlying relational structures of (process) data have gained severe attraction. Thus, a swift towards a network-centric perspectives took place. Consequently, different fields of science contributed to data collection and the statistical analysis of such data. 

As a result, tremendous developments have taken place and several different network models have been derived. Most contributions have been applied in social network analysis including exponential random graph models also known as ERGMs \citep{Koskinen:Lusher:Robins:2011}, multilevel network models \citep{Lazega2016} as well as agent-based models \citep{SnijdersEtAl2010}.  For a detailed presentation of social network models in general we refer the interested reader to  \cite{Carrington2005}. Besides social network models, network models have extensively been studied within statistical physics. Most prominently, these network models include the Watts-Strogatz small-world model (cf. \cite{Watts:1999}). Extensive reviews of both prominent fields of network models are provided by \cite{Goldenberg:EtAl:2010} as well as \cite{Kolaczyk:2009}.

Within the network analysis framework, only a very small proportion of papers have dealt with relational structures with respect to spatial data. \cite{VerHoef:Peterson:Theobald2006} and \cite{VerHoef:Peterson:2010} discussed the usage of directed networks to capture transition flows occurring in the spatial domain. Even less work has been made restricted to the subfamily of planar point processes. In planar point processes one is interested in the analysis of randomly occurring observations within a bounded area. Most commonly, the relational structures taken into account with respect to planar point process data are restricted to univariate - possibly marked - point processes in the spatial domain.

One family of such models for planar point processes was presented by \cite{Marchette:2004}. \cite{Penrose:Yukich:2001} and \cite{Penrose:2003,Penrose:2005} discussed adaptations of random graph models in which a node of a network graph corresponds to a single realization of the process. These random graphs are also known as neighbor networks and include spanning trees or tessallations graphs as special cases. So, neighbor networks are related to geometrical measures and usually restricted to univariate (possibly marked) point processes.

Another family of structural models for planar point processes that is similar in spirit, but different in detail, was presented by \cite{Okabe:Yamada:2001} as well as \cite{Ang:2010} and \cite{Ang:Baddeley:Nair2012} who derived an extension of Ripleys' $K$-function \citep{ripley:76} to planar point processes on linear networks.  \cite{Baddeley:Jammalamadaka:Nair:2014} recently proposed an extension for planar multitype processes on linear networks. Additional details of and further contributions to this field of structural modeling within this linear network formalism can be found in  \cite{Okabe:Satoh:2009, Okabe:Sugihara:2012} and \cite{Borruso:2005,Borruso:2008}. Generally, these models focus on realizations of point processes that occur randomly on a linear network in a bounded region on some planar space. The analyses of such processes can thus be regarded as a generalization of the well-established planar point process methodology. Consequently, the treatment of point process data on linear networks in the spatial domain requires refined versions of classical spatial statistics derived for planar point processes. Inevitably, these refinements are needed in order to incorporate the specific characteristics and geometry of the underlying graph structures. Examples of events that might occur on linear networks are crimes committed in the streets of a specific district, road accidents on a traffic network or insect specifies on a mortar network on walls.

Although several statistics for point processes on linear networks have been derived, certain limitations due to the linear network formalism remain and should be emphasized. First, these models are restricted to linearity of the network and most commonly directionless line segments. Second, the first- and second-order moments and related statistics are defined with respect to given radii over line segments which might result in biased estimates of complex structures such as in the case of events that occur as clusters on street segments. Such complex pattern would have only been recognized if they felt into a given radius, otherwise this information would not be considered in estimation.

\begin{figure}[h!]
\begin{center}
\begin{subfigure}{.29\textwidth}
\includegraphics{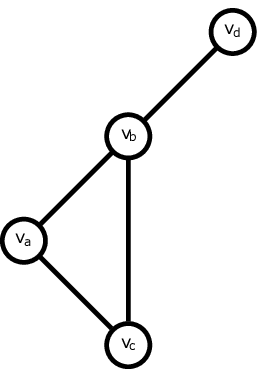}
\caption{}\label{Fig:1a}
\end{subfigure}
\begin{subfigure}{.29\textwidth}
\includegraphics{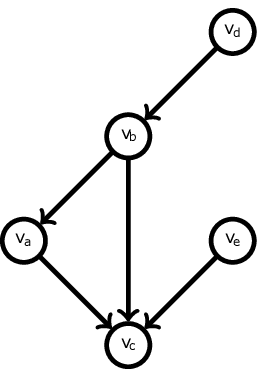}
\caption{}\label{Fig:1b}
\end{subfigure}
\begin{subfigure}{.29\textwidth}
\includegraphics{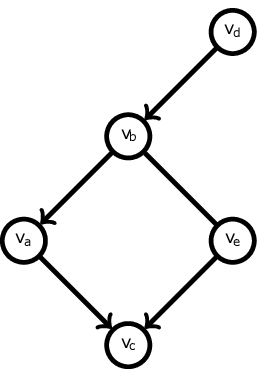}
\caption{}\label{Fig:1c}
\end{subfigure}
\caption{Examples of simple graphs: a) an undirected graph, b) a digraph and c) a simple mixed graph}
\label{Fig:1}
\end{center}
\end{figure}

To address these methodological drawbacks we introduce an alternative formulation of networks with respect to planar point processes adopting some well-known ideas of graphical modeling. To this end, we define different intensity  measures related to the number of events that occur on edges joining pairs of nodes and neighboring vertices as well as vertex-edge sequences in form of paths and trails for different types of graphs. Here, our approach includes directed graphs as well as partially directed graphs besides directionless network graphs. Simple examples of all three types of graphs as discussed in this paper are shown in Figure \ref{Fig:1}. Figure \ref{Fig:1a} displays a simple undirected graph based on four nodes. A visualization of a directed graph is given in Figure \ref{Fig:1b}. Lastly, Figure \ref{Fig:1c} shows a graph build on partially directed edges which can be seen as a combination of Figure \ref{Fig:1a} and \ref{Fig:1b}.   

The different configurations of the network as shown in Figure \ref{Fig:1}
will give us the opportunity to control for different forms of possible transitions in a given network. Directed network graphs are a suitable choice when movements along a network are directional restricted. Possible questions to be answered might be the occurrence of particles in rivers under consideration of the flowing direction. In addition to this, partially directed network graphs will be of interest whenever transition flows are only partially directional restricted such as in case of one-way streets within a road map.

The idea to use general graphical modeling formalism to analyze structural relations in planar point processes has recently been presented by \cite{Eckardt:2016} in form of undirected graphs. Focussing on the underlying conditional dependence structure in multitype planar point patterns, an undirected graphical model was introduced which captures the relation between component processes. Here, each component process is represented as a node and edges are associated with partial statistics defined in the frequency domain. Differently, the vertex set as used here corresponds to unique locations of the network and events are assumed to appear on edges defined as intervals of arbitrary length joining pairs of vertices. This leads to an alternative formulation for possibly multitype point processes that occur on arbitrary networks. To illustrate the types of networks that are considered in this paper, we shall discuss a simple example.  

\subsection{Motivating Example}

\begin{figure}[h!]
\begin{center}
\includegraphics[scale=1.2]{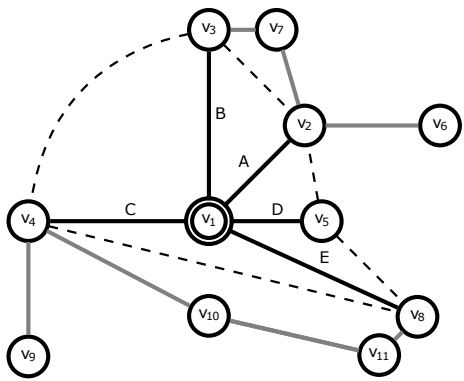}
\caption{Artificial irregularly shaped graph. Neighborhood of node $v_1$ indicated by dashed lines. Solid black lines indicate edges belonging to the neighborhood of $v_1$. Solid gray lines indicate edges not directly connected to $v_1$.}
\label{Fig:2}
\end{center}
\end{figure}

For motivation, we assume that one is interested in analyzing point patterns on irregular road networks. Such a road network might be displayed in Figure \ref{Fig:2} where the edges correspond to streets and the vertices correspond to sectioning elements such as crossings. Suppose that we are interested in the analysis of point patterns that are related to position $v_1$ as highlighted by a double circle in the plot. For $v_1$ we immediately see that the area covered by the set of nodes which are directly connected to $v_1$ is neither circular nor regular in terms of distances. Namely, the set of nodes contains $v_2$ to $v_5$ and  $v_8$. 

Obviously, one can ask several questions when moving along the network departing from $v_1$. With respect to accidents, one might be interested in the intensity for each possible choice of road $A$ to road $E$. Additionally, one might be interested in the mean intensity of accidents in adjacent roads or in the intensity when moving along a sequence of street segments. So, several different intensities could be of interest and generated from the network graph.     

Even more complex, some roads of the traffic network might also be non-linear or directional restricted. Especially in directional restricted or partly directional restricted networks, characteristics of events with respect to impacts of nodes which run square or descent from a certain node could offer important information.
All these restrictions are not captured by circular formulations and remain concealed focusing on only one intensity measure. In contrast, e.g. for $v_1$ a mean intensity with respect to the adjacent street sections controls for any type of irregularity of the network. 

In this paper we aim at presenting an alternative approach with respect to the analysis of multitype planar point patterns that occur on networks of arbitrary shape. The \texttt{R}-code has been made available as supplementary material. The plan of the paper is the following. Section 2 presents the methodological approach of dealing with point patterns occurring on a network. We adapt graph theory to different types of networks and define summary statistics for such cases. An application to crime data is analyzed in Section 3. The paper ends with some final conclusions.

\section{Spatial point patterns on networks}

This section presents two alternative approaches of dealing with point process data which occur on a network. The methodological fundamentals of point processes which appear on linear networks will be presented in Section  \ref{Sec:lpp}. A brief introduction of central graph theoretical concepts is given in Section  \ref{Sec:graph}. Lastly, Section \ref{Sec:AlternativeModel} presents an alternative approach to model such data and relates this to different types of network graphs. 

\subsection{Processes on linear networks}\label{Sec:lpp}

Consider a point pattern $x$ on a linear network $\Lnet$ as a realization of a point process $X$ on the linear network space $\Lnet\subset\mathbb{R}$. Here, the $\Lnet$-space appears as the union over a finite set of lines segments $l_i$. A line segment between two planar points $u$ and $v$ is defined as $l(\cdot)=\lbrace tu+(1-t)v:0\leq t\leq 1\rbrace$ where $|l_i\cap l_j|=0$ for all $l_i\neq l_j\in\Lnet$. An important concept for linear networks is the shortest path distance $d_\Lnet(u,v)$. For two points $u$ and $v$ in the $\Lnet$-space, $d_\Lnet(u,v)$ is the length of the shortest path along the network between $u$ and $v$ and is set to infinity in case that $v$ is not reachable from $u$. Obviously, if $v$ is not reachable from $u$ then $u$ is not reachable from $v$ due to the underlying undirected graph structure.      
The set of points whose shortest path distance lies within radius $r$ with center point $u$ in $\Lnet$ is called the disc $b_\Lnet(u,r)$,
\begin{equation}
b_\Lnet(u,r)=\lbrace v\in \Lnet:d_\Lnet(u,v)\leq r \rbrace
\end{equation}
The relative version, $\partial b_\Lnet(u,r)$, expresses the set of points on $\Lnet$ lying exactly $r$ units away from $u$. As a counting measure with respect to $\partial b_\Lnet(u,r)$, the circumference $m_\Lnet$ encodes the number of points of $\Lnet$ lying exactly $r$ units away by the shortest path from $u$.

As introduced by \cite{Okabe:Yamada:2001}, replacing the Euclidean distance $ \Vert x_i-x_j  \Vert$ by the shortest path distance $d_\Lnet(x_i,x_j)$ we can define a $\Lnet$-version of Ripleys' $K$-function as follows   
\begin{equation}\label{def:net.K}
\hat{K}(r)=\frac{|\Lnet|}{n(n-1)}\sum^n_{i=1}\sum_{j \neq i}\mathbbm{1}_{\lbrace d_\Lnet(x_i,x_j)\leq r\rbrace}
\end{equation}
which expresses the expected number of events that fall into a certain network distance at radius $r$ of a fixed point. Additionally,
the pair correlation function on a linear network is given as
\begin{equation}\label{def:net.pcf}
g_\Lnet(r)=\frac{1}{\sum_i 1\slash \hat{\lambda}(x_i)}\sum^n_{i=1}\sum_{j \neq i}\frac{\kappa(d_\Lnet(x_i,x_j)-r)}{\hat{\lambda}(x_i)\hat{\lambda}(x_j)m_\Lnet(x_i,d_\Lnet(x_i,x_j))}
\end{equation}
where $\kappa(\cdot)$ is a kernel function and $\lambda(\cdot)$ denotes the intensity function of a point pattern.

\subsection{Graph preliminaries}\label{Sec:graph}

To describe the three types of simple graphs as shown in Figure \ref{Fig:1} formally, we give some basic notation and introduce the necessary graph terminology needed in Section  \ref{Sec:AlternativeModel}. Here, we consider a graph $G$ as a tuple consisting of a finite set of vertices $V=\lbrace v_1,\dots,v_k\rbrace$ and a finite set of edges 
 $E\subseteq V\times V$ joining the vertices where $E(G)\cap V(G)=\emptyset$. We say that a pair of vertices $(v_i,v_j)$ is \textit{adjacent} if $(v_i,v_j)\in E( G)$. Otherwise, $(v_i,v_j)$ are \textit{non-adjacent}. We only consider loopless graphs built on simple edges such that neither $(v_i, v_i)\in E( G)$ nor multiple edges are joining a pair of nodes.
 
Obviously, the shape of the edges may vary leading to different types of graphs. We remark that the following terminology is restricted to undirected as well as directed edges. An \textit{undirected} or \textit{unoriented} edge exists if the pairs $(v_i, v_j)$ and $(v_j,v_i)$ are both in the edge set $E(G)$ given $v_i\neq v_j$.  Undirected edges will be indicated by writing $v_j \mrf v_i$. One important notion to specify adjacency in case of undirected edges is given by the \textit{neighborhood}, defined as $\nach\left(v_j\right)=\lbrace
v_i: v_j\mrf v_i\rbrace$.

In contrast, an edge is called \textit{directed} or \textit{oriented} if the ordered pair $(v_i, v_j)\in E( G),~v_i\neq v_j$. Thus, only $(v_i, v_j)\in E( G)$ while $(v_j,v_i)\notin E( G)$. Directed edges will be denoted by $v_i\bnr v_j$. Similarly to the undirected case, adjacency between vertices may be described in terms of different  sets of relational structures such as \textit{parents} ($\pa(\cdot)$) or \textit{children} ($\child(\cdot)$). Formally, theses sets are given as $\pa(v_j)=\lbrace v_i: v_i\bnr v_j\rbrace$ and  $\child(v_i)=\lbrace v_j: v_i\bnr v_j\rbrace$. Additionally, two nodes form the \textit{co-parents} $\cop(v_i)=\lbrace v_j: \child(v_j)\cap\child(v_i)\neq\emptyset\rbrace$ whenever they share a common child.

\begin{figure}
\begin{center}
\includegraphics*[width=.55\textwidth]{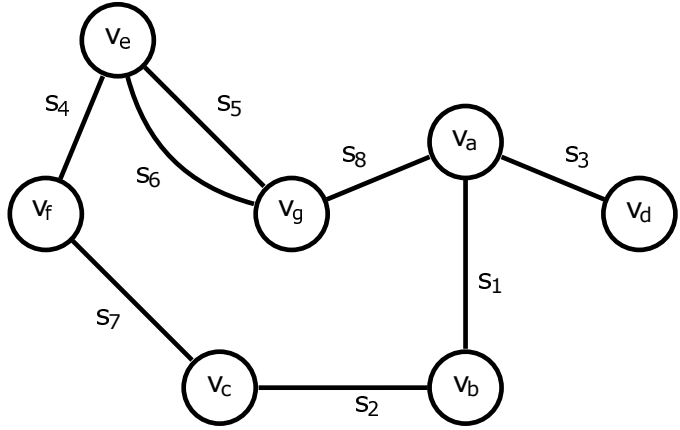}
\caption{Possible vertex edge sequences in undirected multigraphs}
\label{fig:3}
\end{center}
\end{figure}

Taking several edges into account we can define different forms of movements along sequences of vertices and edges in a graph. Formally, we  
treat $\left(v_0,e_1,v_1,e_2,\ldots,v_{k-1},e_k,v_k\right)$ as a sequence of vertices and edges of $ G$ with endpoints $v_0$ and $v_k$ such that  $\forall~e_i, 1\leq i\leq k $ the pair $v_{i-1}$ and $v_{i}$ is joined by $e_i$. This sequence of potentially repeating pairs of vertices is a \textit{walk} of \textit{length} $k$ in $G$. Introducing restrictions on walks leads to specific forms of vertex-edge movements which will be of interest in Section  \ref{Sec:AlternativeModel}. A walk in which any edge of a sequence is traversed at most once is a \textit{trail} or route. Further restrictions lead to a \textit{path}, which is a trail that passes through every node of a sequence exactly once. A path from $v_i$ to $v_j$ will be denoted by $\pi_{ij}$. A path with identical endpoints is a \textit{cycle} and a cycle of length one is a loop. Examples of a walk, a trail and a path are shown in Figure \ref{fig:3}.

Here, the sequence $(v_e,s_5,v_g,s_5,v_e,s_4,v_f,s_7,v_c)$ is a walk since $s_5$ as well as $v_e$ are traversed twice. Omitting any edge which occurs more than once yields a trail. Thus, the node-edge list $(v_e,s_5,v_g,s_6,v_e,s_4,v_f,s_7,v_c)$ forms a trail in $ G$. Lastly, a path is present for example as $\pi_{af}=(v_a,s_1,v_b,s_2,v_c,s_7,v_f)$.

Naturally, a generalization of the former terminology leads to the definition of undirected graphs and directed graphs (digraphs) as shown in Figure \ref{Fig:1a} and Figure \ref{Fig:1b}. A graph is called an \textit{undirected graph} if all edges in $E(G)$ are undirected. Consequently, a \textit{digraph} is a graph exclusively build on directed edges. Thus, a \textit{directed acyclic graph} (DAG) is a digraph without any directed cycles. A simple graph without any partially directed cycle  build on directed as well as undirected edges is called a \textit{partially directed graph}. Here, $ V( G)$ is partitioned into $k$ \textit{blocks} $\mathcal{T}$ such that  $ V(G)= \mathcal{T}_1\cup \mathcal{T}_2\cup\ldots\cup \mathcal{T}_{k-1}\cup \mathcal{T}_k$ and 
\begin{itemize}
\item[i.] $v_i\bnr v_j$ only if $v_i\in \mathcal{T}_i,v_j\in \mathcal{T}_j, i<j$
\item[ii.] $v_i\mrf v_j$ only if $v_i\in \mathcal{T}_i,v_j\in \mathcal{T}_i$.
\end{itemize}   
All statistics introduced in this paper will be defined for undirected graphs and extended for directed as well as partially directed graphs.
 
In Section  \ref{Sec:AlternativeModel} we explicitly use the number of adjacent nodes which will be expressed by means of the degree $\dg(\cdot)$. Depending on the graph taken into account, different calculations of the degree appear. For an undirected graph the minimal degree $\dg_{min}(G)$ and the maximum degree $\dg_{max}(G)$ of a graph are given as
\[
\dg_{min}(G)=\min(\dg(v_i)|v_i\in V(G))
\]
and
\[
\dg_{max}(G)=\max(\dg(v_i)| v_i\in V(G)).
\] 
These formulae are slightly modified in case of directed graphs such that the minimum and maximum degree are obtained as
\[
\dg^-(G)=\sum_{i\in\pa(j)}v_i
\]
and
\[
\dg^+(G)=\sum_{i\in\child(j)}v_i.
\]  
Again, the degree measures in the context of partially directed graphs follow as a combination of different degree measures. Thus, the complete degree of vertex $v_i, v_i \in V(G)$ is given as
\[
\dg_{cg}(G)=\sum_{i\in\nach(j)\cup\family(j)}v_i
\] 
where $\family(v_i)=\lbrace \child(v_i)\cup \pa(v_i)\rbrace$.

\subsection{Alternative formulation}\label{Sec:AlternativeModel}

Similar to Section  \ref{Sec:lpp}, we consider realizations of a possibly marked point pattern that appear on a network. Oppositely, there is no need to consider the point process to be simple.

 More formally, we treat a network as a graph $G=(V,E)$ where events randomly occur on arbitrary locations over an edge joining two vertices. Let $V(s_v)$ denote the set of vertices of a spatial network graph $G$ in which every element $v_i(s_{v_i})\in V(s_v)$ is indexed with a pair of fixed coordinates $s_{v_i}=(x_{v_i},y_{v_i})$. Thus, the $i$-th node contained in $V$ represents a certain position of interest in the network, e.g. the $i$-th road crossing.  Then, the edges can be thought of as intervals of arbitrary length that connect two locations of interest represented as nodes in space. The realizations of a point pattern that fall into specific edges can then be regarded as a random collection of points that fall into a specific interval spanned between two locations in a planar space.

We now concentrate on the definition of such intervals defined over edges. To start, let $\mathcal{S}_{E(G)}=\lbrace s_{e_1},\ldots,s_{e_k}\rbrace$ denote the set of $k$ edge intervals $s_{e_i}=(s_{v_i},s_{v_j})$ between any pair of vertices $(v_i,v_j)$ contained in $V( G)$. For a point process $X(\tilde{s})$, let $\tilde{s}=(\tilde{x}, \tilde{y})$ be the location within a closed interval belonging to $\mathcal{S}_{E(G)}$. Here, we note that we treat $s_{e_i}=(s_{v_i},s_{v_j})$ as fixed and $\tilde{s}=(\tilde{x}, \tilde{y})$ as random. Extensions to multiple types of disjoint events appear naturally.  

In this sense, the interval is spanned between two vertices of a graph and a sequence of such intervals forms a path. So, for any two vertices which are connected by a path we can calculate the distance based on the length of the   path. Here, the length of a path on a network graph is the number of the intervals (edges) between two nodes. From this, we obtain the distance $d_G(v_i,v_j)$ between two vertices of the network graph as the length of the shortest path. Differently to $d_\Lnet(u,v)$, and the related statistics as briefly described in Section  \ref{Sec:lpp}, these distances do not assume regularity or a certain degree of circularity. As in case of $d_\Lnet(\cdot,\cdot)$, we set any distance  $d_G(\cdot,\cdot)$ between two nodes which are not reachable via a path in a network graph to infinity.

To avoid any circular formulation in form of radii centred around given points, we extend the given graph definitions to differently sized polynomial areas of interest covered by a network. These differently sized polynomial areas are addressed by using a non-negative integer-valued order $\xi_G$.  Thus, for $\xi_G=2$ we could then obtain a refined neighborhood that includes all edges which are reachable from a fixed vertices within a $2$-edges distance. Referring to Figure \ref{Fig:2} this extended definition of $\nach(v_1)$ includes all vertices shown in the plot.  

Based on this set-up, we now introduce counting measures and statistics with respect to points contained in $\mathcal{S}_{E(G)}$ with respect to different types of network graphs. 

\subsubsection{Undirected networks}\label{Sec:UG}
For an undirected network graph, a counting measure appears as
\[
N(s_{e_i})=\sum\mathbbm{1}_{\lbrace x_{v_i}\leq \tilde{x}\leq x_{v_j},y_{v_i}\leq \tilde{y}\leq y_{v_j}\rbrace}X(\tilde{s}), x_{v_i}<x_{v_j},y_{v_i}<y_{v_j}.
\]
Thus, $N(s_{e_i})$ expresses the number of points that fall into the edges, and are used to define different intensity measures related to undirected networks. 

To start, a pairwise intensity function follows as 
\[
\lambda(s_{e_i})=\lim_{|ds_{e_i}|\rightarrow 0}\left\{\frac{\mathbbm{E}\left[N(ds_{e_i})\right]}{|ds_{e_i}|}\right\}, s_{e_i}\in\mathcal{S}_{E(G)}
\] 
where $ds_{e_i}$ is an infinitesimal interval. A more local intensity can additionally be defined in terms of the neighborhood of a vertex. So, setting $ e_i=(v_i,v_j)$ this yields an intensity measure for vertex $v_i$ with regard to the neighboring vertices 
\[
\lambda(v_i)=\frac{1}{|\dg(v_i)|}\sum_{v_j\in\nach(v_i))}\lambda(s_{e_i}). 
\]

Besides these local measures, we can define alternative intensity measures with respect to movements along a network graph. For a path $\pi_{ij}$ from $v_i$ to $v_j$, we can define a path-wise intensity measure
\[
\lambda(\pi_{ij})=\frac{1}{|\mathcal{N}_\pi|}\sum_{v_j\in\pi_{ij}}\lambda(s_{e_i})
\]
where $\mathcal{N}_\pi$ is the number of consecutive intervals contained in $\pi_{ij}$.

Apart from first-order characteristics, one can easily extend second-order statistics within our alternative approach. 

A modification of \cite{Okabe:Yamada:2001} $\Lnet$-version of Ripleys' $K$-function substituting  $d_G(v_i,v_j)$ for $d_\Lnet(x_i,x_j)$ in \eqref{def:net.K} yields to
\begin{equation}\label{def:net.K.alternative}
\hat{K}_G(\xi_g)=\frac{|E|}{n(n-1)}\sum^n_{i=1}\sum_{j \neq i}\mathbbm{1}_{\lbrace d_G(v_i,v_j)\leq \xi_g\rbrace},
\end{equation}

where $|E|$ is the number of edges denoted as the size of the network graph.

\subsubsection{Directed networks}

Directed network graphs require a modification of the previously introduced statistics in terms of the direction of the arc, namely parents and children. For parents, we define
\[
N(s_{e_i}^{in})=\sum\mathbbm{1}_{\lbrace\pa(x_{v_i}\leq \tilde{x}\leq x_{v_j},y_{v_i}\leq \tilde{y}\leq y_{v_j})\rbrace}X(\tilde{s}), x_{v_i}<x_{v_j}y_{v_i}<y_{v_j}
\]
as a counting measure which expresses the number of events on an edge leading to a vertex of interest. The opposite relation, the number of events on an edge departing from a vertex of interest yields to a counting measure defined over the set of children nodes, namely
\[
N(s_{e_i}^{out})=\sum\mathbbm{1}_{\lbrace\child(x_{v_i}\leq \tilde{x}\leq x_{v_j},y_{v_i}\leq \tilde{y}\leq y_{v_j})\rbrace}X(\tilde{s}), x_{v_i}<x_{v_j}y_{v_i}<y_{v_j}.
\]

As benefit to the undirected case, we obtain different intensity statistics from the counting measures which contain additional information on the direction of the network flow. The pair-wise intensities are then given as  
\[
\lambda(s_{e_i}^{in})=\lim_{|ds_{e_i}^{in}|\rightarrow 0}\left\{\frac{\mathbbm{E}\left[N(ds_{e_i}^{in})\right]}{|ds_{e_i}^{in}|}\right\}, s_{e_i}^{in}\in\mathcal{S}_{E(G)}
\] 
and
\[
\lambda(s_{e_i}^{out})=\lim_{|ds_{e_i}^{out}|\rightarrow 0}\left\{\frac{\mathbbm{E}\left[N(ds_{e_i}^{out})\right]}{|ds_{e_i}^{out}|}\right\}, s_{e_i}^{out}\in\mathcal{S}_{E(G)}.
\] 

Again, setting $ e_i=(v_i,v_j)$ we obtain an intensity measure for vertex $v_i$ with regard to the parents of $v_i$  as 
\[
\lambda^{in}(v_i)=\frac{1}{|\dg^+(v_i)|}\sum_{v_j\in\pa(v_i))}\lambda(s_{e_i}) 
\]
and the children of $v_i$ as
\[
\lambda^{out}(v_i)=\frac{1}{|\dg^-(v_i)|}\sum_{v_j\in\child(v_i))}\lambda(s_{e_i}).
\]
Extending Ripleys' $K$-function to directed networks results in two different directed versions of \eqref{def:net.K.alternative} based on a generic direction depending Delta function $\delta^{(\star)}$. Here, $\delta^{(\star)}$ could either be related to directed edges or, if $\xi_G\geq 2$, to direction preserving paths pointing to ($\delta^{(<)}$) or departing from ($\delta^{(>)}$) a distinct node. From this, we obtain a generic directed $K$-function $\hat{K}^{(\star)}_G$ as     
\begin{equation}\label{def:net.K.alternative:directed}
\hat{K}^{(\star)}_G(\xi_G)=\frac{|E|}{n(n-1)}\sum^n_{i=1}\sum_{j \neq i}\mathbbm{1}_{\lbrace d_G(v_i,v_j)^{\delta^{(\star)}}\leq \xi_G\rbrace}
\end{equation}
which describes the probability of a further event conditional on a given observation within a direction preserving path of length $\xi_G$.  

\subsubsection{Partially directed networks}

Counting measures and related statistics for partially directed networks appear as combination of the measures introduced for undirected and directed network graphs. Since partially directed network graphs consist of directed as well as undirected edges, all previously introduced counting measures and statistics are in general applicable to partially directed networks. Additional information can be achieved as union over certain sets of edges. An intensity measure related to the union of parents, children and neighbors of a specific node could be defined as
\[
\lambda^{cg}(v_i)=\frac{1}{|\dg^{cg}(v_i)|} \lambda^{out}(v_i)\cup \lambda^{in}(v_i)\cup \lambda(v_i).
\]
Alternative measures appear naturally as re-definition of $\lambda^{cg}(v_i)$ taking only certain unions into account, e.g. $\pa(\cdot)\cup\child(\cdot)$ which contain information about the directed adjacent edges. Measures defined with respect to $\nach(\cdot)\cup\child(\cdot)$ will exclude any edge pointing to a node of interest, while the set $\nach(\cdot)\cup\pa(\cdot)$ will exclude any edge departure from a specific node of interest.  

Again, a partially directed $K$-function appears as a slightly extension of \eqref{def:net.K.alternative:directed} as     
\begin{equation}\label{def:net.K.alternative:chain}
\hat{K}^{(\ast)}_G(\xi_G)=\frac{|E^{\delta^{(\ast)}}|}{n(n-1)}\sum^n_{i=1}\sum_{j \neq i}\mathbbm{1}_{\lbrace d_G(v_i,v_j)^{\delta^{(\ast)}}\leq \xi_G\rbrace}
\end{equation}
where $\delta^{(\ast)}$ could either be related to undirected edges of be defined identical to $\delta^{(\star)}$.
 
\subsubsection{Distributions over network structures and linkages to geostatistical processes}\label{sec:geostat}

The previously introduced graph perspective inherits additional benefits that are highlighted here. These are (a) the possibility to define a conditional distribution over network structures, and (b) to establish a link to geostatistical processes and thus to adopt well-known geostatistical methodologies. 

For (a) we concentrate on the undirected network formalism as introduced in Section  \ref{Sec:UG}. To derive the conditional distribution, we consider the neighborhood intensity measure for a point pattern and note that this measure allows to establish a close relation to the seminal paper of \cite{Besag:74} and its results with respect to the conditional distribution of lattice data. Similarly to \cite{Besag:74}, we can define the conditional distribution over the nodes of the network graph as 
\[
v_i|\nach(v_i)\stackrel{d}{\sim}f_{v_i|\nach(v_i)}(\lambda(v_i)).
\]
Obviously, a generalization of the expression to higher dimensional patterns appear naturally.  

With respect to (b), we note that the vertices of the network graph are associated within our formalism as presented here with the segmenting locations of the network taken into account. These locations are indexed with pairs of coordinates and are treated as fixed. By definition, all intensity measures are calculated with respect to a specific fragment or set of the network graph, e.g. edges, neighbors or parents. Treating for example the neighborhood intensities as vertex attributes, the intensity measures appear as continuous measure recorded for fixed locations over a network. Thus, we can establish a linkage from point patterns to geostatistics using our alternative network formalism.

Similarly, the edges can be thought of as a countable list of unique identification numbers related to fixed edges. Edge intensity can then be addressed as a  continuous measures recorded at a fixed position in space.

\subsubsection{Spatio-temporal extensions}

We now extend our approach to situations where additionally temporal information is available. Thus, we are concern with realizations of a possible marked point process $X(\tilde{s},\tau_x)$ where again $\tilde{s}=(\tilde{x}, \tilde{y})$ denotes the random location within a closed interval belonging to $\mathcal{S}_{E(G)}$ and $\tau_x\in T \subset \mathbbm{R}_+$ denotes the time of occurrence of an event in $\tilde{s}$. Here, instead of continuous time we assume that $\tau_x$ is discretized.  To this end, the temporal evolution of a spatio-temporal pattern appears as an ordered list $t_0<t_1<\ldots <t_n$ of length $n$ taking values in $\mathbbm{N}_0$. In addition, we introduce a $\delta$-function related to the $n$ time slices. Precisely, for each interval $\left(t_i,t_{i+1}\right], i=0,\ldots,n-1$ we define $\delta_i, i=1,\ldots,n$  such that
\[
\delta_i=
\begin{cases}
1\text{~if~} t_i \leq \tau_{x} < t_{i+1}\\
0\text{~else}.
\end{cases}
\]

This yields a modified counting measure for undirected networks as follows
\[
N(s_{e_i},\tau_i)=\sum\mathbbm{1}_{\lbrace x_{v_i}\leq \tilde{x}\leq x_{v_j},y_{v_i}\leq \tilde{y}\leq y_{v_j}\rbrace}X(\tilde{s})^{\delta_i}, x_{v_i}<x_{v_j},y_{v_i}<y_{v_j},
\]
which expresses the number of points that fall into an edge interval within a given time slice. The number of events that occurred in a network segment up to time $t$ will then be of the form
\[
N(s_{e_i},t)=\prod_{\tau_i< t}^\tau N(s_{e_i},\tau_i).
\]

\section{Application: crime-related data}

\subsection{Data and network description}
 
The data used as example here reports georeferenced coordinates of phone calls received by the police station in the city of Castellon (Spain) from January 1$^{st}$, 2012 to December 30$^{th}$, 2013.  The listed calls were received at the local police call centre or transferred by 112 emergency service to the local police call centre. Geo-codification was  performed indirectly by local officials based on precise address information provided by the caller. Following this procedure, it yields a spatio-temporal dataset where each entry contains the spatial coordinates and the time of occurrence of the call. The calls comprise up to nine different types of crimes or anti-social categories. See Table \ref{tab:tabtwo} for their description.

The city of Castellon is divided into $108$ census sub-areas with an overall surface of $108659 km^2$. According to the information given by the city hall, the total amount of inhabitants is $181616$ of people at the end of 2010. Here, the analysis is based on a subset of phone calls received from the city center that has an overall surface of $8616 Km^2$ divided in $89$ census sub-areas and 130294 inhabitants.

\begin{figure}[H]
\begin{center}
\includegraphics[scale=0.5]{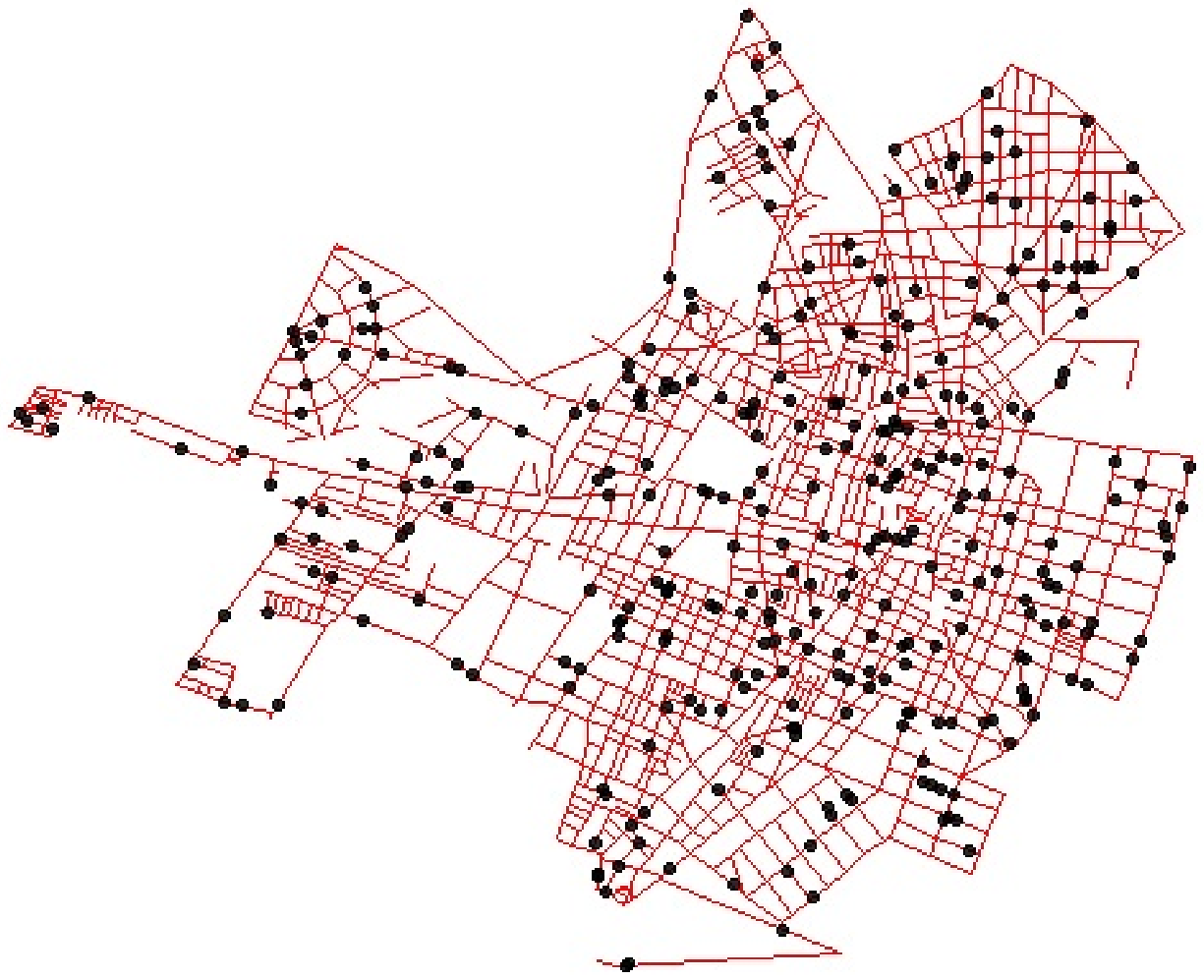}
\end{center}
\caption{Castellon traffic network where solid pink lines indicate streets and events are plotted as black dots   \label{fig:4}}
\end{figure}

For the analysis, we selected in total 1611 segmenting locations of the traffic network treated as the vertex set of our network graph. So, the vertex set contains 1611 single nodes of which two were isolated. To each location we attribute the precise georeferenced coordinates. For any edge in the edge set we calculated the interval length as the squared geodesic distance between pairs of these coordinated vertices. The corresponding traffic network and the recorded events indicated as black dots are shown in Figure \ref{fig:4}.

\begin{figure}
\begin{center}
\includegraphics[scale=0.5]{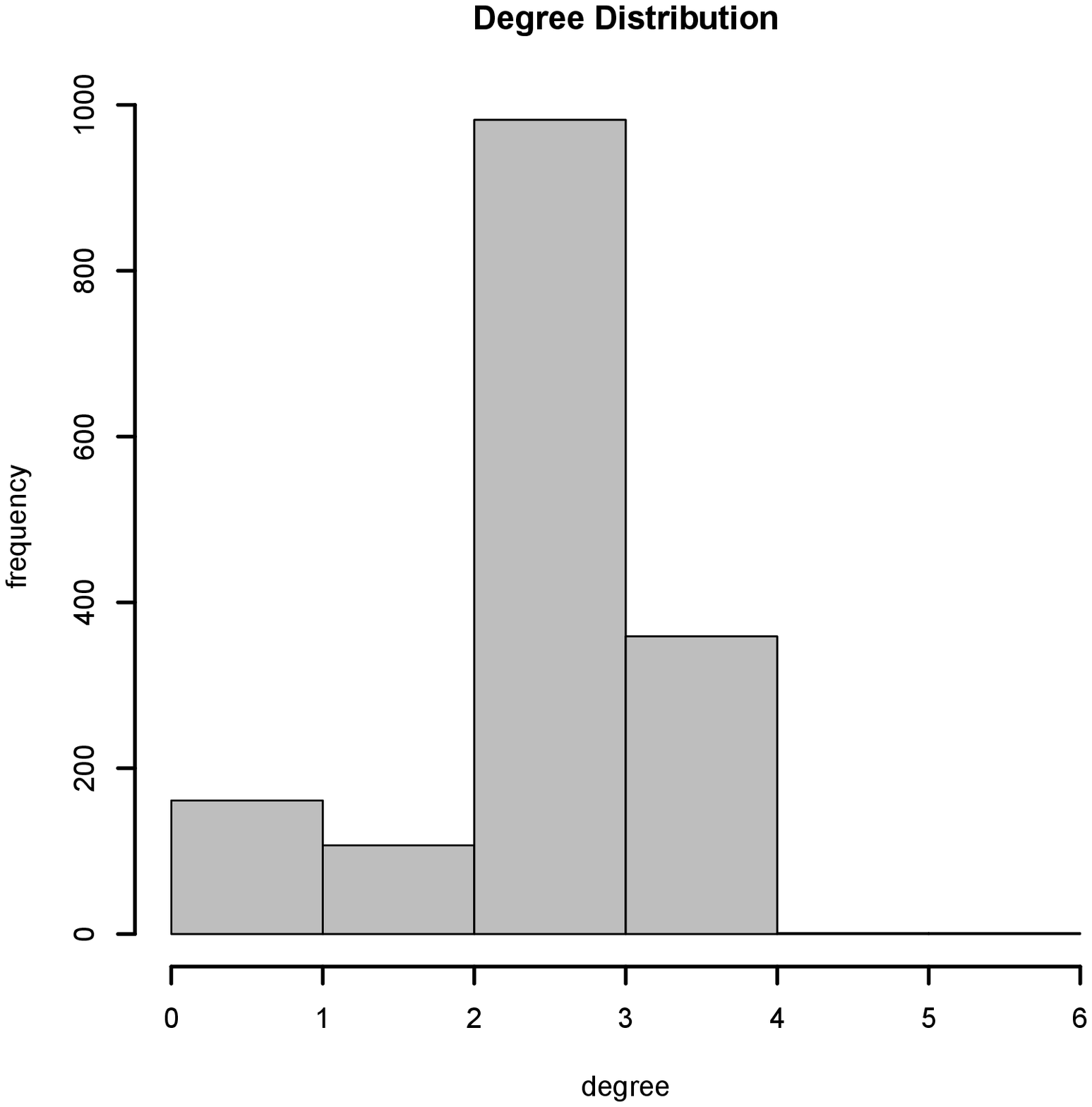}
\end{center}
\caption{Degree distribution of the Castellon road network \label{fig:5}}
\end{figure}

Generally, we considered the network graph to be directionless. A first impression of the shape of the network is given by the degree distribution as shown in Figure \ref{fig:5}.  

Here, the mean degree is 2.96 and at most 6 edges are incident to a single node in the network graph. 

\subsection{Spatial patterns on the network}

The summary statistics for the edgewise as well as neighborhood intensities for overall process are reported in Table \ref{tab:tabone}. Generally, both measures indicated low probabilities of becoming affected to crime within the streets of Castellon. However, certain areas as well as street segment showed a higher edgewise as well as neighborhood intensity, namely 1.95. Thus, the observed pattern is considered to be heterogeneous.

\begin{table}
\caption{Intensity measures for the Castellon traffic net   \label{tab:tabone}}
\begin{center}
\begin{tabular}{rrrrrrr}
Intensity Measure & Min. & 1st Qu.   &  Median  &  Mean    & 3rd Qu.  &  Max.\\\hline
edgewise & 0.00309 & 0.03617 & 0.08178 & 0.13640 & 0.17740 & 1.94800 \\
neighborhood & 0.00123 & 0.01255 & 0.02601 & 0.05218 & 0.05838 & 1.94800 
\end{tabular}
\end{center}
\end{table}

Additionally, we considered the observed pattern as multitype pattern where the marks correspond to the categorization of the phone calls. To this end, we computed edgewise intensities for 9 different categories including violence and robbery, accidents and emergency assistance, drugs, under age, unhealthy wastes, fear perception, vehicles, animal and lastly others. The calculated edge intensities are reported in  Table \ref{tab:tabtwo}.

\begin{table}
\caption{Edge intensity measures for 9 different crime types committed on the Castellon traffic net. VR = violence and robbery, AE = accidents and emergency assistance, DR = drugs, UA = under age, UW = unhealthy wastes, FP = fear perception, VE = vehicles, AN = animal, OT = others. \label{tab:tabtwo}}
\begin{center}
\begin{tabular}{rrrrrrr}
Category & Min. & 1st Qu.   &  Median  &  Mean    & 3rd Qu.  &  Max.\\\hline
VR & 0.00259 & 0.01533  & 0.02827 & 0.04451 & 0.05766 & 0.43230 \\ 
AE & 0.00447 & 0.02250  & 0.04181 & 0.08496 & 0.08496 & 1.30700 \\
DR & 0.00357 & 0.01108  & 0.01648 & 0.02388 & 0.02692 & 0.17290 \\
UA & 0.00279 & 0.00824  & 0.01136 & 0.01743 & 0.01867 & 0.09047 \\
UW & 0.00557 & 0.01088  & 0.01149 & 0.01285 & 0.01579 & 0.02314 \\
FP & 0.00259 & 0.01117  & 0.01906 & 0.02872 & 0.03284 & 0.32740 \\
VE & 0.00188 & 0.00888  & 0.01297 & 0.01860 & 0.01981 & 0.14190 \\
AN & 0.00265 & 0.01110  & 0.01689 & 0.02469 & 0.02939 & 0.21610 \\
OT & 0.00174 & 0.00937  & 0.01311 & 0.01616 & 0.01902 & 0.06539 \\
\end{tabular}
\end{center}
\end{table}

Again, we observe slightly variation over the distribution of the nine categories. As for the overall process, the probability of becoming a victim of crime in nearly zero in all categories. In contrast, especially the upper 25 percent and maximum of the distribution show a slightly different picture. While the edgewise intensity of under age, unhealthy wastes and others remains below 10 percent, Table \ref{tab:tabtwo} displays a moderate increase in the intensities of violence and robbery as well as of fear perception. Additionally, the intensity of accidents and emergency assistance ranges from a minimum of 0.00 to a maximum of 1.31. These observations mean, that while driving or walking along certain road segments the probability of becoming a victim of violence and robbery chances from zero to nearly 0.5. This would mean, that every second person who is moving along a high intensity street segment might be affected by violence or robbery. This argumentation is in line with the heterogeneous edge intensities of fear perception which might express a relation of the presence of violence and robbery to the individual safety feelings.          

\subsection{Smoothing and clustering of multitype networks}

In addition to the intensity calculations as previously discussed, we considered the neighborhood intensity measures for all nine categories taken into account as vertex attributes to our network graph. Thus, the node list is associated with an intensity matrix of dimension $1611\times 9$. 

\begin{figure}[H]
\begin{center}
\includegraphics[scale=0.5]{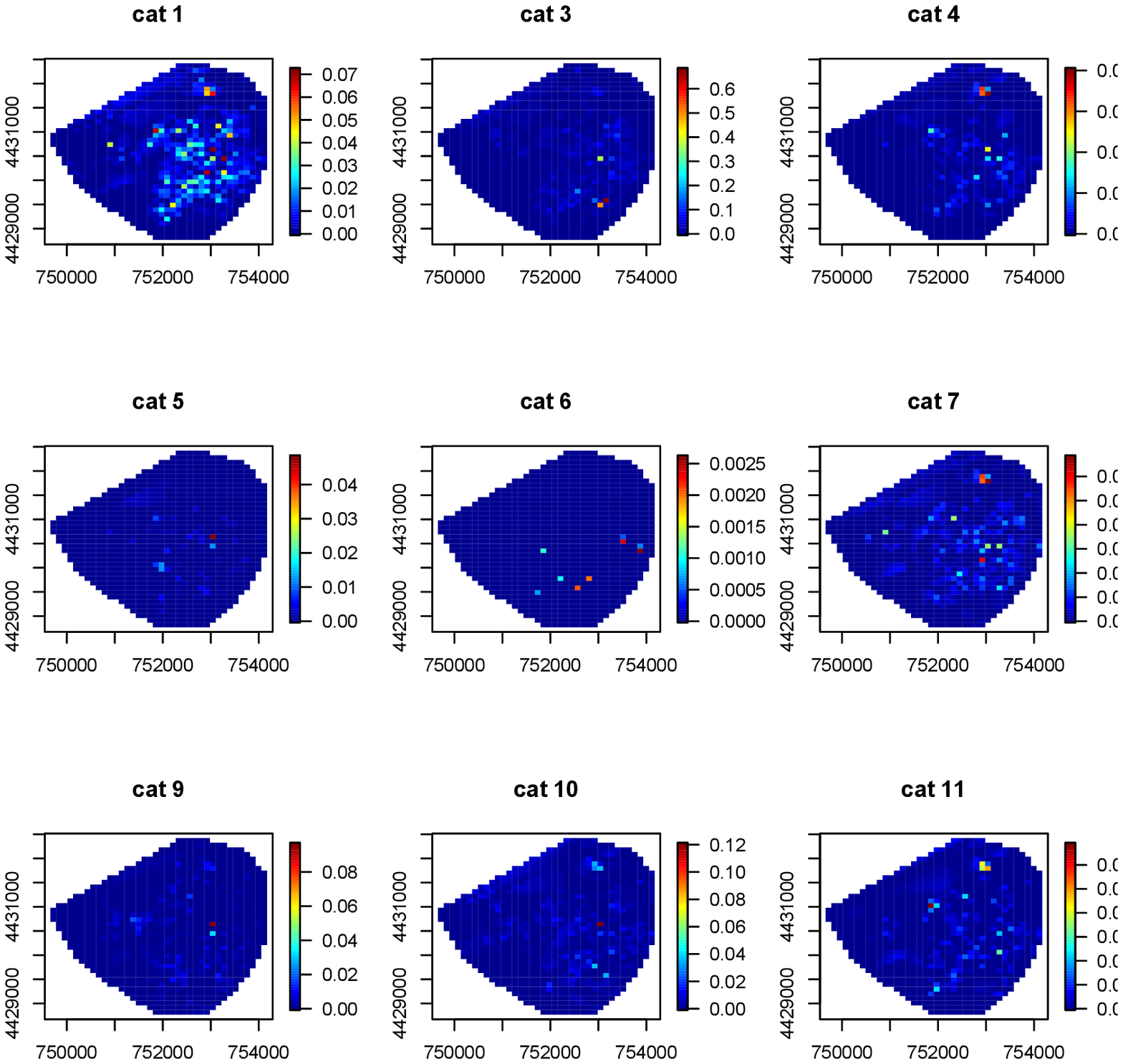}
\end{center}
\caption{Smoothed neighborhood intensities of the Castellon road network: CAT1 = violence and robbery, CAT3 = accidents and emergency assistance, CAT4 = drugs, CAT5 = under age, CAT6 = unhealthy wastes, CAT7 = fear perception,  CAT9 = vehicles, CAT10 = animal, CAT11 = other \label{fig:6}}
\end{figure}

Based on this matrix we conducted two additional analysis of the node-wise crime structure. 
First, as described in Section  \ref{sec:geostat}, we treated the neighborhood intensities of all nine categories separately as continuous values associated to fixed locations on the network. For each crime category, a geostatistical smoother was used to detect different crime regions within the road map of Castellon. The results are displayed in Figure \ref{fig:6}.

Here, we observe varying intensities and varying ranges between the different crime categories. The strongest fluctuation appears for violence and robbery, although the intensities only range from 0.00 to 0.07. In contrast, the nodal mean intensities for accidents and emergency assistance with a maximum value of 0.6 shows a less fluctuating behavior. Here, high values only emerge in the southern areas of the city border. This result matches with the location of a main avenue surrounding the city of Castellon. 

\begin{figure}[H]
\begin{center}
\includegraphics[scale=0.5]{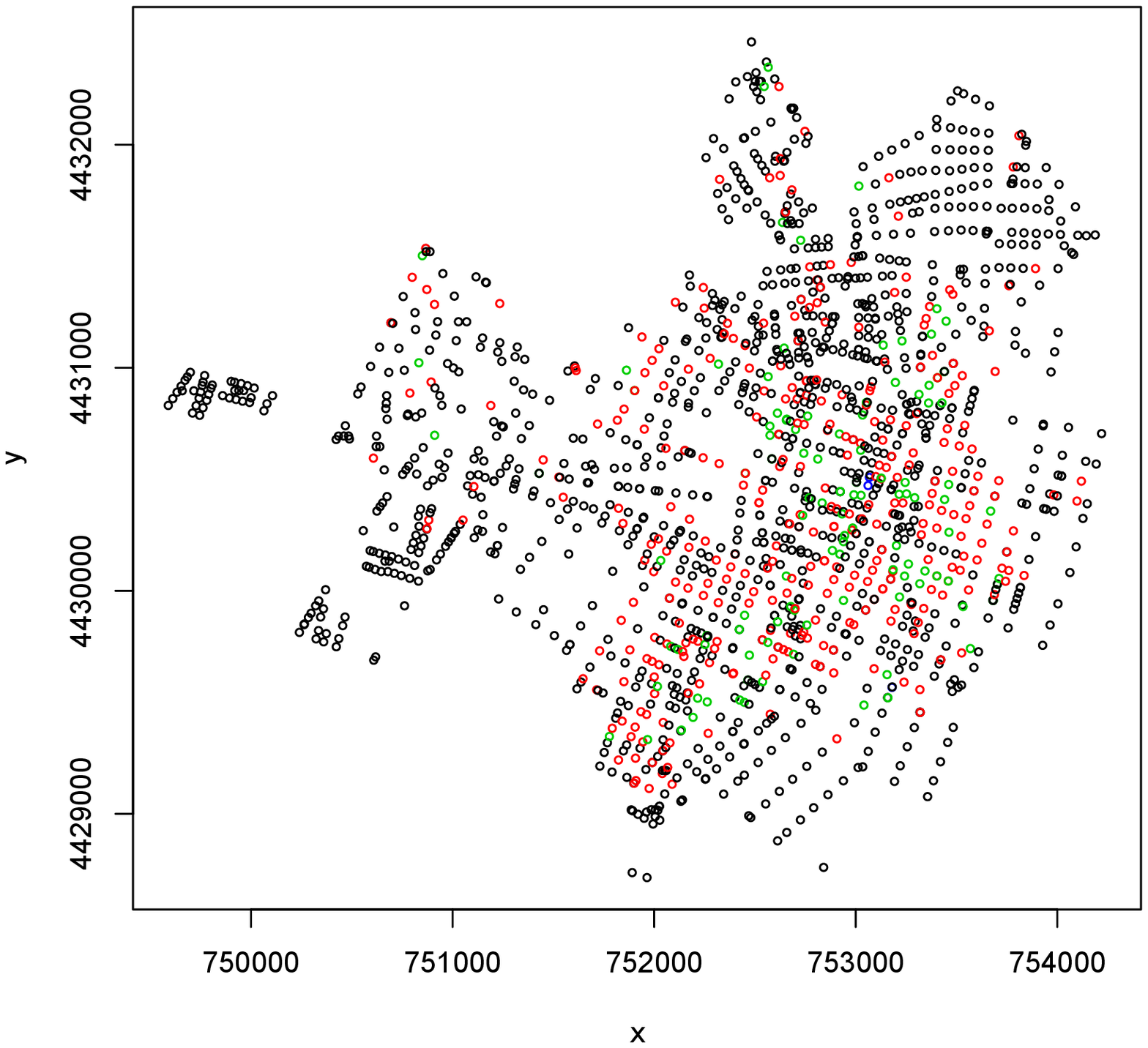}
\end{center}
\caption{Crime clusters of the Castellon road network. The colors blue, red, green and black express distinct cluster components resulting from hierachical clustering of all nine crime categories attribute to vertices of the traffic net  \label{fig:7}}
\end{figure}

In addition, we computed clusters over all nine category specific neighborhood intensities based on the complete $1611\times 9$ dimensional matrix using hierarchical clustering and the Ward algorithm. As a result, we obtained four cluster components which were then used as mark attributed to the fixed vertices of the traffic net. The corresponding marked traffic net is depicted in Figure \ref{fig:7}. In this plot, a clear structure is shown where the blue colored cluster component only slightly appears in the city centre.  

\subsection{Spatio-temporal network analysis}

For illustration, we also considered the univariate spatio-temporal point pattern  which was recorded for the complete observation window from January 1$^st$, 2012 to December, 31$^th$, 2013. For this data, we calculated weekly differences related to the reference time  January 1$^{st}$, 2012. Based on these calculations, we divided the selected 104-week observation window into 12 consecutive sub-windows based on 8-week time-slices. This results in 11 8-week intervals and one 4 week period (weeks 100 - 104).   

\begin{figure}[H]
\begin{center}
\includegraphics[scale=0.5]{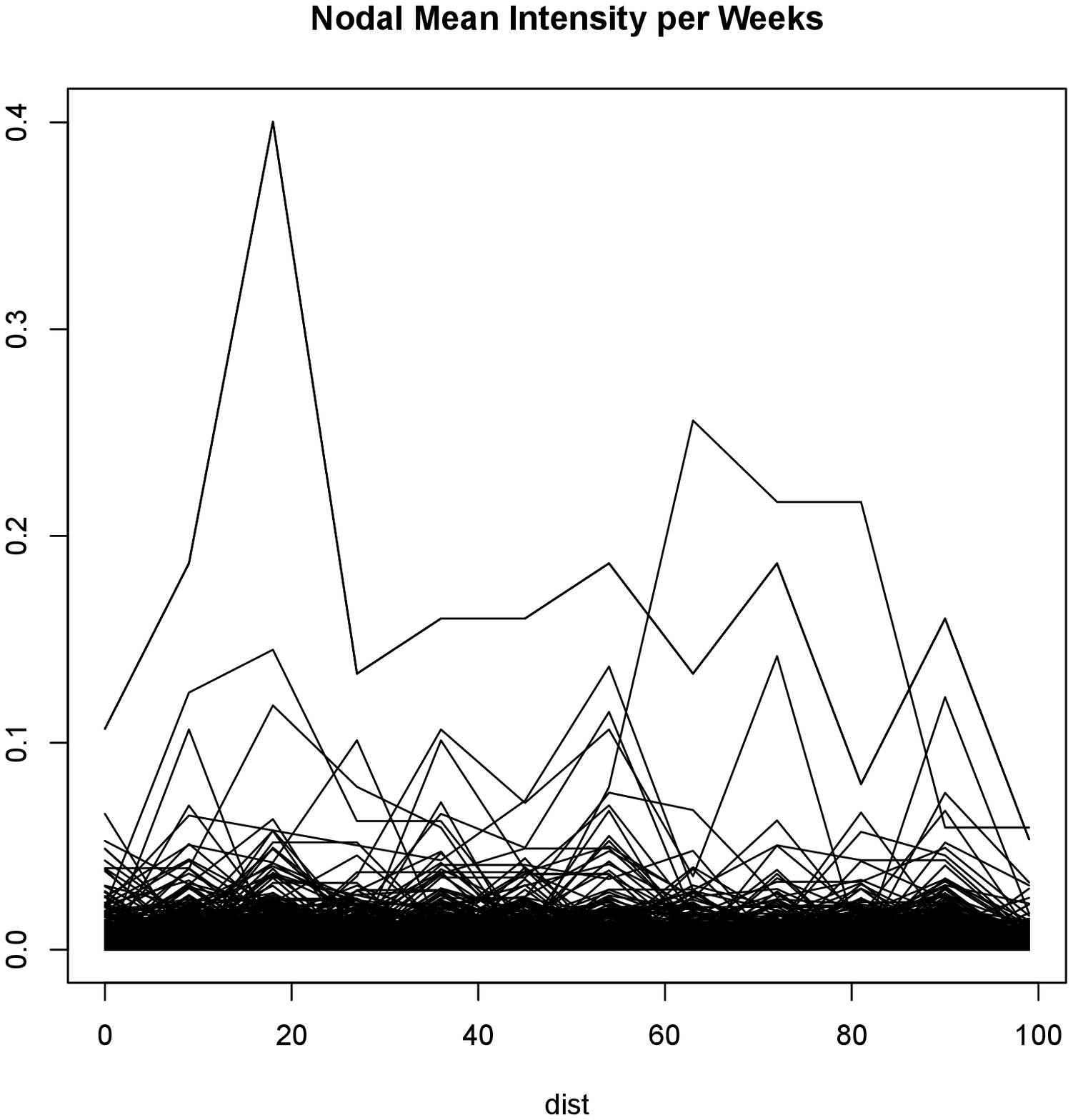}
\end{center}
\caption{Node-wise mean crime intensity for Castellon calculated separately for 12 consecutive temporal intervals of a 104 week observation window \label{fig:8}}
\end{figure}

In a first step, we calculated the nodal mean intensities for each time-slice. These intensities were then attributed as a temporally ordered sequence to the vertices of the traffic net such that one can gain insights into the temporal evolution of the nodal mean crime intensities. This evolution for the complete 104-week observation window is displayed in Figure \ref{fig:8}.

Here, for the highest values of the subset of non-zero nodal mean intensities we observe a clearly cyclic behavior with high peaks in the early weeks of 2012 and 2013.  
%

\section{Discussion}
This paper has introduced an alternative approach for modeling point patterns related to network data. Different from existing considerations, several alternative intensity measures and versions of Ripleys' $K$-function for undirected, directed and partially directed networks have been introduced which offer previously undetected information. All these definitions are related to different sets of vertices which allows to calculate the point process statistics independently of given radii centred at any vertex of interest.  

Different from the linear network formalism, the proposed methodology considers events that occur on a distinct segment of a network regardless of the length of the interval. This allows to control for all types of point patterns such as clustered point pattern which might be present on an arbitrary shaped street segment. Especially non-linear network segments can be included in form of sums over piecewise linear subsegments.

Our proposal connecting graph theory and point processes opens up new ideas of research from both the methodological and the practical points of view. The \texttt{R} code will be provided to help new researchers in getting new insights into this framework. 

\bibliographystyle{Chicago}
\bibliography{network}
\end{document}